\documentclass[sigconf]{acmart}

\AtBeginDocument{%
  }

\usepackage{amsfonts}
\usepackage{multirow}
\usepackage{array}

\newcommand{\pct}{\%}
\newcommand{\x}{\times}

\copyrightyear{2026}
\acmYear{2026}
\setcopyright{cc}
\setcctype{by}
\acmConference[MMAsia '26]{ACM Multimedia Asia}{December 15--18, 2026}{Hanoi, AA, Vietnam}
\acmBooktitle{ACM Multimedia Asia (MMAsia '26), December 15--18, 2026, Hanoi, AA, Vietnam}
\acmDOI{10.1145/3862739.3863388}
\acmISBN{979-8-4007-3133-4/2026/12}

\begin{document}

%% ---- Title ---------------------------------------------------------------
\title[Temporal Stabilization for Robust Delay Estimation]{Learning Temporal Stabilization for Double-Talk-Robust Delay Estimation in Acoustic Echo Cancellation}

%% ---- Authors --------------------------------------------------------------
%% Double-anonymous review: the `anonymous' class option suppresses all
%% author information. Restore the author block below at camera-ready.
\author{Lu Ma}
\affiliation{%
 \institution{JD.com, Inc.}
 \city{Beijing}
 \country{China}}
\email{iamroad@163.com}

\author{Yongliang Wang}
\affiliation{%
 \institution{JD.com, Inc.}
 \city{Beijing}
 \country{China}}
\email{wangyongliang1@jd.com}

\author{Zhewen Su}
\affiliation{%
 \institution{JD.com, Inc.}
 \city{Beijing}
 \country{China}}
\email{suzhewen.1@jd.com}

\renewcommand{\shortauthors}{Lu Ma et al.}

%% ---- Abstract --------------------------------------------------------------
\begin{abstract}
Acoustic echo cancellation (AEC) depends on accurate time-delay estimation (TDE): the adaptive canceller converges only after the bulk delay between the far-end reference and the microphone is compensated. GCC-PHAT remains the standard TDE front-end and, in our stress tests, wins or ties against a fully learned estimator under interference, low SNR, strong reverberation, loudspeaker nonlinearity, and noisy references. Its principal failure mode is short-window sustained double-talk, where near-end speech creates spurious correlation peaks that classical smoothing and confidence gating cannot reliably suppress. We therefore retain GCC-PHAT as a fixed, non-learnable backbone and learn only a 13.2K-parameter causal 2-D CNN stabilizer over the time--lag correlogram. The stabilizer exploits a simple temporal distinction: true delays form persistent ridges, whereas double-talk artifacts appear as transient spikes. Across synthetic and real AEC-Challenge echo paths, it reduces MAE and gross-outlier rates by approximately $3$--$5\x$ over tuned classical estimators and the WebRTC AEC3 matched filter. A faithful AEC3-style downstream testbed further shows that these delay gains translate into lower echo-leak rates and faster post-jump ERLE recovery in time-varying regimes, closing much of the gap to the oracle-delay ceiling. An optional coarse-to-fine pre-alignment stage extends the same residual stabilizer to stable 100--400\,ms bulk delays without retraining, while keeping the learned component focused on short-window double-talk robustness.
\end{abstract}

%% ---- CCS Concepts -----------------------------------------------------------
%% TODO(authors): regenerate this block at https://dl.acm.org/ccs/ccs.cfm
%% and replace the IDs below with the generated ones.



\maketitle

%% ============================================================================
\section{Introduction}
\label{sec:intro}
Bulk-delay compensation is a prerequisite for AEC: delay errors of tens of milliseconds reduce echo-return loss enhancement (ERLE) and may prevent the adaptive filter from locking~\cite{benesty2001,haykin2014}. GCC-PHAT~\cite{knapp1976} is therefore widely used as the TDE front-end. Prior work has explored DNNs built on GCC-like or phase-based features, primarily for source localization and direction-of-arrival estimation~\cite{chakrabarty2019,kowalk2022,pertila2019}, but these methods typically optimize clean-condition accuracy rather than the real-time failure modes encountered in AEC.

This paper does not propose another replacement estimator. Instead, we first identify the one regime in which GCC-PHAT is genuinely fragile---short-window sustained double-talk---and then address that regime while preserving the classical backbone. A causal 2-D CNN is trained only to stabilize the time--lag correlogram, leaving the GCC-PHAT estimate itself untouched. Larger device or network delays are treated separately: bulk alignment is handled by a non-learned coarse stage rather than by enlarging the learned window. Finally, we test whether improved delay estimates matter downstream by coupling each estimator to a faithful AEC3-style adaptive canceller.

Our contributions are fourfold: (1)~a robustness analysis that isolates GCC-PHAT's principal weakness and the limited reach of classical fixes (Section~\ref{sec:sweep}); (2)~a fixed-backbone, learned-stabilizer design motivated by the failure of learnable front-ends to preserve broadband phase coherence (Section~\ref{sec:method}), together with an optional coarse-to-fine pre-alignment stage for stable large bulk delays (Section~\ref{subsec:large_delay}); (3)~paired delay-accuracy evaluation against tuned classical baselines and the deployed AEC3 matched filter on synthetic and real echo paths, including the 100--400\,ms coarse-to-fine regime (Section~\ref{sec:delay}); and (4)~a downstream AEC evaluation showing that delay quality becomes ERLE gains precisely in time-varying double-talk regimes (Section~\ref{sec:downstream}).

%% ============================================================================
\section{Where GCC-PHAT Fails}
\label{sec:sweep}
We stress-tested a purely learned estimator against GCC-PHAT across AEC-relevant conditions. In these preliminary sweeps, GCC-PHAT wins or ties under small stable delays, microphone-side interference, low SNR, strong reverberation, loudspeaker nonlinearity, and noisy references; the learned estimator reaches only 100--200\,ms MAE where GCC-PHAT achieves approximately 0--3\,ms (Section~\ref{sec:method}). GCC-PHAT's weakness appears only when the analysis window is short enough to bound algorithmic latency and the near-end talker remains active. Under this regime, near-end speech can produce correlation peaks that are as sharp and tall as the true echo peak, causing occasional catastrophic delay jumps.

Table~\ref{tab:opening} quantifies this failure and the limits of standard remedies such as causal median filtering and coherence- or sharpness-based gating~\cite{gansler1996}. A causal median removes roughly half of the damage, while a peak-sharpness gate helps little because the spurious near-end peak is also sharp. The residual problem is therefore not peak detection but temporal reliability: deciding whether a strong peak is a persistent delay or a short-lived double-talk artifact.

\begin{table}[t]
\centering\footnotesize
\caption{The principal GCC-PHAT weakness (100\,ms window, sustained double-talk, true GT). The PSR gate abstains on low-confidence frames; its MAE and outlier rate are therefore not directly comparable and are omitted.}
\label{tab:opening}
\begin{tabular}{lccc}
\toprule
Method & MAE (ms) & tol@5 & outlier \%$>$25 \\
\midrule
Raw GCC-PHAT (argmax) & 21.5 & 0.58 & 38\pct \\
\ + causal median      & 11.7 & 0.74 & 20\pct \\
\ + PSR confidence gate & --  & 0.64 & -- \\
\bottomrule
\end{tabular}
\end{table}

%% ============================================================================
\section{Method}
\label{sec:method}

\begin{figure*}[t]
\centering
\includegraphics[width=0.94\textwidth]{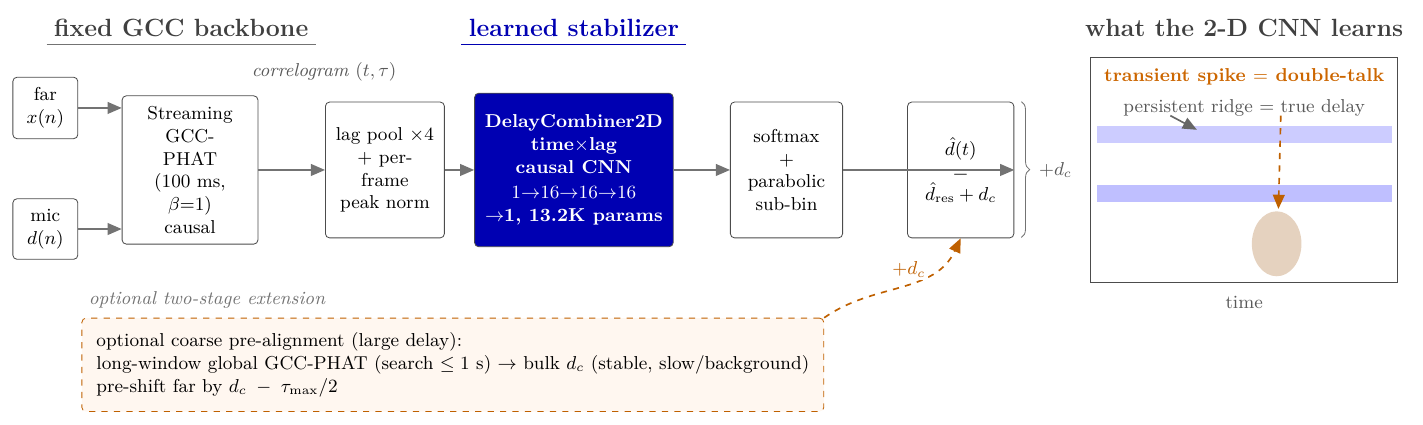}
\caption{System overview. A fixed streaming GCC-PHAT backbone produces a causal time--lag correlogram, which is pooled, peak-normalized, and passed to a lightweight 2-D causal CNN stabilizer. The stabilizer learns to preserve persistent delay ridges while suppressing transient double-talk spikes. An optional coarse pre-alignment stage handles large delays outside the 80\,ms residual search range.}
\Description{Block diagram of the delay-stabilization system. Far-end and microphone signals enter a fixed streaming GCC-PHAT module, followed by lag pooling and per-frame peak normalization. A learned DelayCombiner2D network processes the time-lag correlogram and outputs the stabilized delay through softmax and parabolic sub-bin refinement. A dashed optional coarse pre-alignment block provides a bulk delay offset for large delays. An inset contrasts a persistent true-delay ridge with a transient double-talk spike.}
\label{fig:arch}
\end{figure*}

\noindent\textbf{Design principle.} An earlier learnable-front-end variant, inspired by raw-waveform front-ends such as SincNet~\cite{ravanelli2018}, failed to reach competitive accuracy, with MAE of 100--200\,ms compared with approximately 0--3\,ms for pure GCC. The underlying issue is that a learned front-end disrupts the broadband phase coherence on which GCC-PHAT depends. We therefore keep the GCC-PHAT backbone fixed and non-learnable, and introduce learning only after the correlogram has been formed.

\noindent\textbf{Pipeline.} As shown in Figure~\ref{fig:arch}, streaming GCC-PHAT (100\,ms window, 1024-sample hop, $\beta{=}1$, causal left padding) produces a correlogram $(B,T,\tau{+}1)$ with $\tau_{\max}{=}80$\,ms. Lag-axis average pooling ($\x4$, 0.25\,ms/bin) and per-frame peak normalization feed \textbf{DelayCombiner2D}, a causal time$\times$lag CNN ($1{\to}16{\to}16{\to}16{\to}1$, lag kernels $11/9/7/5$, time kernel $3$, FrameGroupNorm, LeakyReLU). Softmax over lag bins followed by parabolic sub-bin refinement yields the stabilized delay. All 13.2K learnable parameters reside in the combiner; the GCC backbone has none.

\noindent\textbf{Temporal inductive bias.} The network learns a distinction that per-frame peak statistics cannot express: a true delay persists as a ridge across time, whereas a double-talk artifact is a transient spike. Causality is strict because GroupNorm is applied independently to each frame. An optional reliability head was trained with binary cross-entropy but is omitted from the champion model because the ablation shows it is unnecessary for accuracy.

\noindent\textbf{Training.} Training uses fully synthetic echo paths with real speech, delays of 0--80\,ms, sustained double-talk, delay jumps, and music, singing, or noise interference. We use weight EMA, ReduceLROnPlateau, cross-entropy over lag bins, and smooth-$L_1$ loss on the sub-bin delay. Evaluation is paired: every method sees identical frames on a held-out split.

\subsection{Coarse-to-Fine Extension for Large Delays}
\label{subsec:large_delay}
The stabilizer operates on an 80\,ms residual window. Larger device or network delays are handled by an optional coarse-to-fine stage rather than by enlarging the learned search space. A long-window global GCC-PHAT search over lags up to 1\,s estimates a slow bulk delay $d_c$. The far-end reference is pre-shifted by $d_c-\tau_{\max}/2$, placing the remaining delay near the center of the residual window, and the final estimate is $\hat d(t)=d_c+\hat d_{\mathrm{res}}(t)$. The coarse stage is slow and runs in the background: it adds no per-frame algorithmic latency and is updated only under hysteresis, while the residual CNN retains short-window double-talk stabilization and fast residual refinement. This division matches the physical assumption that device bulk delay is approximately constant within a clip and extends operation to the 100--400\,ms range with no additional training. A windowed slow-update variant that tracked bulk delay inside each clip was rejected because sustained double-talk could corrupt the bulk estimate; the reported configuration therefore uses stable coarse estimates and leaves short-term variation to the residual stabilizer.

%% ============================================================================
\section{Delay-Accuracy Experiments}
\label{sec:delay}
All baselines share the same fixed GCC front-end so that differences isolate the stabilization stage. We report MAE, tol@5, gross-outlier rate ($|e|{>}25$\,ms), and post-jump recovery. Classical competitors include whitening sweeps ($\beta{=}0.7/0.5$), ML/HT optimal weighting $\psi_{\mathrm{HT}}(f){=}|\gamma(f)|^2/[|G_{xy}(f)|(1{-}|\gamma(f)|^2)]$~\cite{knapp1976}, and a faithful reimplementation of the WebRTC AEC3 matched filter~\cite{webrtc}. A small step size ($\mu{\approx}0.05$) is required for the latter because NLMS does not whiten colored speech~\cite{haykin2014}.

\begin{table*}[t]
\centering\footnotesize
\caption{Delay accuracy against tuned classical baselines and the deployed AEC3 matched filter (same frames, same GT; recov.\ denotes post-jump recovery in analysis windows). Ours improves every variant by approximately $3$--$5\x$; the AEC3 matched filter is competitive for stable delays but degrades under frequent jumps because its adaptive convergence time exceeds the jump interval.}
\label{tab:delay}
\begin{tabular}{lcccccccc}
\toprule
& \multicolumn{3}{c}{Constant delay + sustained DT} & & \multicolumn{4}{c}{Short window + frequent jumps}\\
\cmidrule(lr){2-4}\cmidrule(lr){6-9}
Method & MAE (ms) & tol@5 & out.\%$>$25 & & MAE (ms) & tol@5 & out.\%$>$25 & recov. \\
\midrule
Raw GCC ($\beta{=}1$)      & 14.9 & 0.621 & 25.8\pct & & 15.7 & 0.618 & 27.1\pct & 3.2 \\
GCC $\beta{=}0.7$          & 12.8 & 0.636 & 21.9\pct & & 13.5 & 0.622 & 23.2\pct & 2.9 \\
GCC $\beta{=}0.5$          & 13.8 & 0.581 & 23.4\pct & & 14.7 & 0.562 & 24.0\pct & 3.4 \\
GCC + PSR gate             & 13.0 & 0.704 & 22.5\pct & & 13.4 & 0.696 & 22.8\pct & 4.7 \\
GCC-HT (ML), best classical& 10.1 & 0.720 & 17.3\pct & & 10.9 & 0.680 & 18.1\pct & 3.1 \\
AEC3 matched filter        &  9.8 & 0.693 & 15.3\pct & & 20.5 & 0.372 & 30.4\pct & 15.9 \\
\textbf{Ours (stabilizer)} & \textbf{2.0} & \textbf{0.946} & \textbf{2.7\pct} & & \textbf{3.9} & \textbf{0.892} & \textbf{6.1\pct} & \textbf{3.3} \\
\bottomrule
\end{tabular}
\end{table*}

The learned stabilizer improves both stable and jump regimes (Table~\ref{tab:delay}), and the gain is largest where the task is hardest: under frequent jumps, the AEC3 matched filter collapses to 20.5\,ms MAE and 30.4\pct\ gross outliers, while the proposed stabilizer remains at 3.9\,ms and 6.1\pct.

\noindent\textbf{Real echo paths.} We evaluate two AEC-Challenge protocols~\cite{aecchallenge,cutler2021}. In Route~1A, real far-end loopback drives a synthetic echo path with exact GT ($n{=}64$, 2348 double-talk frames): Ours achieves 4.1\,ms MAE, 0.879 tol@5, and 6.0\pct\ outliers, versus 12.2\,ms, 0.692, and 20.4\pct\ for the best classical baseline. In Route~1B, a real \texttt{farend\_singletalk} echo path is combined with injected near-end speech at SER${=}0$\,dB and scored against gated long-window pseudo-GT ($n{=}64$, 6691 frames): Ours achieves 5.3\,ms, 0.831, and 7.7\pct, versus 14.7\,ms, 0.507, and 26.6\pct. The approximately $3\x$ advantage therefore transfers from synthetic to real echo paths, and Route~1A's exact GT mitigates concerns about pseudo-GT circularity in Route~1B.

\begin{table}[t]
\centering\footnotesize
\caption{Ablation on the held-out split. The temporal 2-D axis is the operative mechanism; the reliability head is unnecessary for accuracy.}
\label{tab:abl}
\begin{tabular}{lccc}
\toprule
Variant & MAE (ms) & tol@5 & outlier \%$>$25 \\
\midrule
A: 2-D + reliability head   & 4.7 & 0.867 & 7.3\pct \\
B: 1-D per-frame + head     & 14.0 & 0.625 & 23\pct \\
C: 2-D, no head (champion)  & 4.7 & 0.880 & 7.4\pct \\
\bottomrule
\end{tabular}
\end{table}

Table~\ref{tab:abl} confirms the source of the gain. The 1-D per-frame variant falls back to the behavior of a classical confidence gate, whereas the 2-D model learns ``persistent ridge versus transient spike'' along time. Training with jumps also removes the usual recovery--stability trade-off: a persistent new ridge indicates a genuine delay change and triggers rapid re-locking, while a transient spike remains suppressed.

\noindent\textbf{Large-delay coarse-to-fine accuracy.} We evaluate the optional coarse-to-fine configuration on constant delays of 100--200, 200--300, and 300--400\,ms under sustained double-talk, with approximately 1800 frames per range. The coarse stage supplies the bulk offset, while the same residual CNN trained only on 0--80\,ms resolves the remaining delay; no additional training or architecture change is used. Table~\ref{tab:large_delay} shows that accuracy degrades gradually with delay range---MAE rises from 3.1 to 4.6\,ms and tol@5 decreases from 0.913 to 0.870---but remains substantially better than GCC-HT throughout. This supports the coarse-to-fine split for stable bulk offsets; it does not claim learned generalization to arbitrary large in-window delays.

\begin{table}[t]
\centering\footnotesize
\setlength{\tabcolsep}{4pt}
\caption{Large-delay coarse-to-fine accuracy under sustained double-talk (constant delay per clip, $n{\approx}1800$ frames per range). The learned residual model is unchanged.}
\label{tab:large_delay}
\begin{tabular}{@{}llccc@{}}
\toprule
Range (ms) & Method & MAE (ms) & tol@5 & out.\%$>$25 \\
\midrule
100--200 & GCC-HT & 7.0 & 0.727 & 14.3\pct \\
         & \textbf{Ours} & \textbf{3.1} & \textbf{0.913} & \textbf{6.8\pct} \\
\addlinespace[2pt]
200--300 & GCC-HT & 8.1 & 0.692 & 16.7\pct \\
         & \textbf{Ours} & \textbf{3.8} & \textbf{0.892} & \textbf{8.2\pct} \\
\addlinespace[2pt]
300--400 & GCC-HT & 8.8 & 0.664 & 17.9\pct \\
         & \textbf{Ours} & \textbf{4.6} & \textbf{0.870} & \textbf{10.0\pct} \\
\bottomrule
\end{tabular}
\end{table}

%% ============================================================================
\section{Downstream Value for AEC}
\label{sec:downstream}
Delay accuracy alone does not establish downstream value. A closed-form least-squares canceller with per-clip bulk alignment is insensitive to frame-level delay jitter, so it collapses all estimators under stable delay. A practical adaptive canceller separates them through two mechanisms: a finite filter length ($\approx$52\,ms), which places incorrectly delayed echo outside the tap range, and adaptation dynamics, which force re-convergence after delay jumps.

%% NOTE(layout): this figure* block is deliberately placed BEFORE the Testbed
%% paragraph. Double-column floats can only appear at the top of the page
%% FOLLOWING the point where LaTeX encounters them; keeping the block here
%% (while page 3 is being built) is what places Figure 2 at the top of page 4.
\begin{figure*}[t]
\centering
\includegraphics[width=0.93\textwidth]{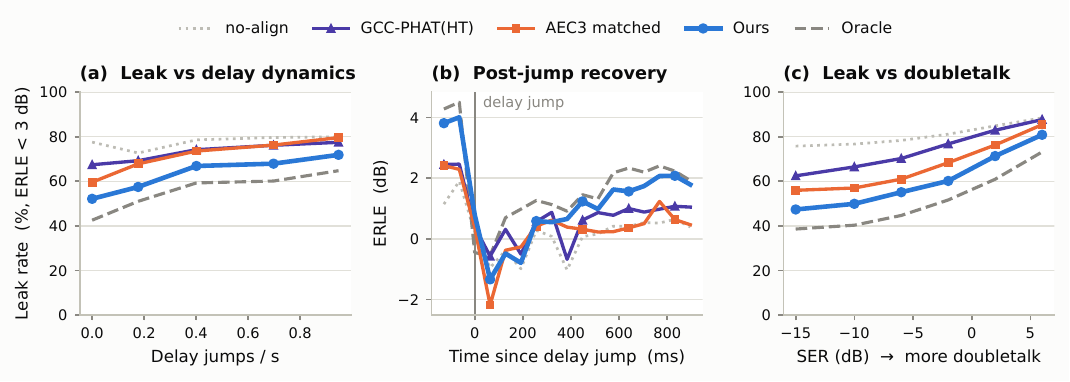}
\caption{Downstream echo cancellation in the time-varying/double-talk niche (isolated-echo ERLE). (a)~Echo-leak rate versus delay-jump frequency; (b)~post-jump ERLE recovery over 77 aligned jumps; (c)~echo-leak rate versus SER. Ours remains closest to the oracle ceiling and below all classical and AEC3 baselines.}
\Description{Three panels reporting downstream echo-cancellation metrics for several delay estimators. Panel a plots echo-leak rate against delay-jump frequency; panel b plots post-jump ERLE recovery; panel c plots echo-leak rate against signal-to-echo ratio. The proposed stabilizer remains closest to the oracle and below classical and AEC3 baselines.}
\label{fig:three}
\end{figure*}

\noindent\textbf{Testbed.} We faithfully reimplemented the core linear stage of AEC3: a partitioned-block frequency-domain adaptive filter~\cite{shynk1992,soo1990} (block 64, FFT 128, 13 partitions, overlap-save), the double-talk-robust regularized Refined Filter Update Gain, refined and coarse shadow sub-filters, and a render-delay buffer driven by the external delay estimate. ERLE is computed on the isolated echo component in steady state. Three implementation details are essential: echo-audible SER with turn-taking near-end silence, alignment headroom for causal filtering, and block-granular render delay quantized to 64-sample blocks.

\noindent\textbf{Mechanism validation.} Table~\ref{tab:mech} contrasts streaming and bulk alignment at a constant 70\,ms delay. Bulk alignment collapses all methods to approximately 6.5\,dB, reproducing the least-squares non-result. Streaming alignment separates them, with the proposed stabilizer approaching oracle performance. The discrimination therefore arises from the finite-length adaptive mechanism rather than an evaluation artifact.

\begin{table}[t]
\centering\footnotesize
\caption{Downstream isolated-echo ERLE (dB), $n{=}48$. Bulk alignment collapses; streaming separates.}
\label{tab:mech}
\begin{tabular}{lccccc}
\toprule
Regime (mode) & no-align & GCC-HT & AEC3-mf & \textbf{Ours} & Oracle \\
\midrule
Const.\ 70\,ms (stream) & $-0.3$ & 1.6 & 3.8 & \textbf{4.2} & 6.3 \\
Const.\ 70\,ms (bulk)   & $-0.3$ & 6.2 & 6.5 & 6.5 & 6.5 \\
Time-varying (stream)   & 3.0 & 3.1 & 2.3 & \textbf{4.2} & 5.2 \\
Large 100--400\,ms (c2f)& $-0.6$ & 3.4 & 2.3 & \textbf{5.1} & 6.0 \\
\bottomrule
\end{tabular}
\end{table}

\noindent\textbf{Large-delay downstream behavior.} The large-delay row in Table~\ref{tab:mech} isolates the same mechanism under coarse-to-fine operation. With stable bulk delays of 100--400\,ms, Ours reaches 5.1\,dB, above GCC-HT (3.4\,dB) and AEC3 (2.3\,dB), and closes most of the gap to the 6.0\,dB oracle. The remaining gap is expected: the coarse stage removes only slowly varying bulk delay and does not change sensitivity to residual jitter or rapid changes.

\noindent\textbf{Tail and event metrics.} Mean ERLE obscures the rare failures that motivate temporal stabilization. Figure~\ref{fig:three} and Tables~\ref{tab:tail}--\ref{tab:sweep} therefore report worst-decile ERLE, echo-leak rate (\% frames with ERLE\,$<$\,3\,dB), and post-jump recovery in the time-varying regime. Ours has the lowest leak rate at every jump frequency and every SER. The AEC3 matched filter approaches no-align performance at 0.95 jumps/s and exhibits the deepest post-jump degradation and slowest recovery.

\begin{table}[t]
\centering\footnotesize
\caption{Time-varying regime: tail and event metrics computed on isolated-echo ERLE.}
\label{tab:tail}
\begin{tabular}{lcccc}
\toprule
Metric & GCC-HT & AEC3-mf & \textbf{Ours} & Oracle \\
\midrule
Mean ERLE (dB)          & 2.20 & 1.94 & \textbf{3.43} & 4.27 \\
p10 (worst-decile) ERLE & $-1.50$ & $-1.77$ & \textbf{$-0.87$} & $-0.12$ \\
Echo-leak rate (\%,$<$3\,dB) & 64.5 & 70.1 & \textbf{52.9} & 46.2 \\
Post-jump recovery (win.)   & 8.4 & 9.7 & \textbf{7.5} & 6.9 \\
\bottomrule
\end{tabular}
\end{table}

\begin{table}[t]
\centering\footnotesize
\caption{Data behind Figure~\ref{fig:three}. Top: echo-leak rate (\%) versus delay-jump frequency. Bottom: echo-leak rate (\%) versus SER (constant delay; higher SER = harder double-talk). Ours remains the lowest non-oracle estimator throughout.}
\label{tab:sweep}
\begin{tabular}{lccccc}
\toprule
Jumps/s & no-align & GCC-HT & AEC3-mf & \textbf{Ours} & Oracle \\
\midrule
0.00 & 78 & 67 & 60 & \textbf{52} & 43 \\
0.18 & 73 & 69 & 68 & \textbf{57} & 51 \\
0.40 & 79 & 74 & 74 & \textbf{67} & 59 \\
0.70 & 80 & 76 & 76 & \textbf{68} & 60 \\
0.95 & 80 & 77 & 80 & \textbf{72} & 65 \\
\midrule
SER (dB) & no-align & GCC-HT & AEC3-mf & \textbf{Ours} & Oracle \\
\midrule
$-15$ & 76 & 62 & 56 & \textbf{47} & 39 \\
$-10$ & 77 & 66 & 57 & \textbf{50} & 40 \\
$-6$  & 78 & 70 & 61 & \textbf{55} & 45 \\
$-2$  & 81 & 77 & 68 & \textbf{60} & 52 \\
$+2$  & 85 & 83 & 76 & \textbf{71} & 61 \\
$+6$  & 88 & 88 & 85 & \textbf{81} & 73 \\
\bottomrule
\end{tabular}
\end{table}

%% ============================================================================
\section{Positioning, Limitations, and Readiness}
\label{sec:limits}
\noindent\textbf{Positioning.} Oracle delay is a ceiling, not a competitor. GCC-PHAT remains near-optimal for clean-reference, small, stable delays, and we do not target the thin margins available there. The contribution is deliberately narrow: a lightweight temporal stabilizer that makes GCC-based delay estimation robust to double-talk and delay variation, with a measurable downstream AEC benefit. It beats its strongest operational competitor, the AEC3 matched filter, precisely in the time-varying regime where adaptive lag aggregation is weakest.

\noindent\textbf{Limitations.} First, the downstream testbed is a faithful reimplementation rather than the actual WebRTC binary, and downstream signals are synthetic; absolute ERLE is low because only the linear stage is modeled, so the supported claim is a relative ranking. Second, the favorable niche---short-window, time-varying, double-talk---can be partly mitigated by mechanisms outside TDE, such as buffer changes, alignment-error tolerance, and nonlinear post-processing; a convincing deployment case would additionally require a real time-varying scenario such as Bluetooth transport or clock drift. Third, learned validation remains limited to the 80\,ms residual window; the 100--400\,ms results rely on a non-learned coarse stage that assumes bulk delay is stable within a clip, so in-segment bulk jumps during sustained double-talk and continuously moving delays remain only partially covered. Running the real WebRTC AEC3 and measuring ERLE on real recordings are the most important next steps.

%% ============================================================================
\section{Conclusion}
\label{sec:conclusion}
Rather than replacing a near-optimal classical estimator, we keep GCC-PHAT fixed and learn only the temporal stabilization it lacks. A 13.2K-parameter causal 2-D CNN reduces delay error and gross outliers by approximately $3$--$5\x$ in short-window double-talk, transfers to real AEC-Challenge echo paths, improves downstream AEC behavior in time-varying regimes, and remains effective for stable 100--400\,ms bulk delays with non-learned coarse pre-alignment. The broader principle: preserve strong classical structure and apply learning selectively to the failure mode it cannot express.

%% ---- Bibliography -----------------------------------------------------------
\clearpage  % keep the bibliography on its own page
\bibliographystyle{ACM-Reference-Format}
\bibliography{reference}

@book{benesty2001,
  author    = {Benesty, Jacob and G{\"a}nsler, Tomas and Morgan, Dennis R. and Sondhi, M. Mohan and Gay, Steven L.},
  title     = {Advances in Network and Acoustic Echo Cancellation},
  publisher = {Springer-Verlag},
  address   = {Berlin, Heidelberg},
  year      = {2001},
}

@book{haykin2014,
  author    = {Haykin, Simon},
  title     = {Adaptive Filter Theory},
  edition   = {5},
  publisher = {Pearson},
  address   = {Upper Saddle River, NJ, USA},
  year      = {2014},
}

@article{knapp1976,
  author  = {Knapp, Charles H. and Carter, G. Clifford},
  title   = {The Generalized Correlation Method for Estimation of Time Delay},
  journal = {IEEE Transactions on Acoustics, Speech, and Signal Processing},
  volume  = {24},
  number  = {4},
  pages   = {320--327},
  year    = {1976},
  doi     = {10.1109/TASSP.1976.1162830},
}

@inproceedings{ravanelli2018,
  author    = {Ravanelli, Mirco and Bengio, Yoshua},
  title     = {Speaker Recognition from Raw Waveform with {SincNet}},
  booktitle = {Proceedings of the 2018 IEEE Spoken Language Technology Workshop (SLT)},
  pages     = {1021--1028},
  year      = {2018},
  publisher = {IEEE},
}

@misc{webrtc,
  author       = {{Google}},
  title        = {{WebRTC} Open-Source Project: Audio Processing Module ({AEC3})},
  howpublished = {\url{https://webrtc.googlesource.com/src/+/refs/heads/main/modules/audio_processing/aec3/}},
  year         = {2025},
  note         = {Accessed: 2026-01-10},
}

@inproceedings{aecchallenge,
  author    = {Sridhar, Kusha and Cutler, Ross and Saabas, Ando and Parnamaa, Tanel and Loide, Markus and Gamper, Hannes and Braun, Sebastian and Aichner, Robert and Srinivasan, Sriram},
  title     = {{ICASSP} 2021 Acoustic Echo Cancellation Challenge: Datasets, Testing Framework, and Results},
  booktitle = {Proceedings of the IEEE International Conference on Acoustics, Speech and Signal Processing (ICASSP)},
  pages     = {151--155},
  year      = {2021},
  publisher = {IEEE},
  doi       = {10.1109/ICASSP39728.2021.9413457},
}

@inproceedings{cutler2021,
  author    = {Cutler, Ross and Saabas, Ando and Parnamaa, Tanel and Loide, Markus and Sootla, Sten and Purin, Marju and Gamper, Hannes and Braun, Sebastian and Sorensen, Karsten and Aichner, Robert and Srinivasan, Sriram},
  title     = {{INTERSPEECH} 2021 Acoustic Echo Cancellation Challenge},
  booktitle = {Proceedings of Interspeech 2021},
  pages     = {4748--4752},
  year      = {2021},
  publisher = {ISCA},
  doi       = {10.21437/Interspeech.2021-1870},
}

@article{shynk1992,
  author  = {Shynk, John J.},
  title   = {Frequency-Domain and Multirate Adaptive Filtering},
  journal = {IEEE Signal Processing Magazine},
  volume  = {9},
  number  = {1},
  pages   = {14--37},
  year    = {1992},
  doi     = {10.1109/79.109205},
}

@article{soo1990,
  author  = {Soo, Jia-Sien and Pang, Khee K.},
  title   = {Multidelay Block Frequency Domain Adaptive Filter},
  journal = {IEEE Transactions on Acoustics, Speech, and Signal Processing},
  volume  = {38},
  number  = {2},
  pages   = {373--376},
  year    = {1990},
  doi     = {10.1109/29.103078},
}

@article{gansler1996,
  author  = {G{\"a}nsler, Tomas and Hansson, M. and Ivarsson, C.-J. and Salomonsson, G.},
  title   = {A Double-Talk Detector Based on Coherence},
  journal = {IEEE Transactions on Communications},
  volume  = {44},
  number  = {11},
  pages   = {1421--1427},
  year    = {1996},
  doi     = {10.1109/26.544458},
}

@article{chakrabarty2019,
  author  = {Chakrabarty, Soumitro and Habets, Emanu{\"e}l A. P.},
  title   = {Multi-Speaker {DOA} Estimation Using Deep Convolutional Networks Trained with Noise Signals},
  journal = {IEEE Journal of Selected Topics in Signal Processing},
  volume  = {13},
  number  = {1},
  pages   = {8--21},
  year    = {2019},
  doi     = {10.1109/JSTSP.2019.2901664},
}

@inproceedings{kowalk2022,
  author    = {Kowalk, Uwe and Doclo, Simon and Bitzer, Joerg},
  title     = {Signal-Informed {DNN}-Based {DOA} Estimation Combining an External Microphone and {GCC-PHAT} Features},
  booktitle = {Proceedings of the International Workshop on Acoustic Signal Enhancement (IWAENC)},
  pages     = {1--5},
  year      = {2022},
  publisher = {IEEE},
  doi       = {10.1109/IWAENC53105.2022.9914754},
}

@inproceedings{pertila2019,
  author    = {Pertil{\"a}, Pasi and Parviainen, Mikko},
  title     = {Time Difference of Arrival Estimation of Speech Signals Using Deep Neural Networks with Integrated Time-Frequency Masking},
  booktitle = {Proceedings of the IEEE International Conference on Acoustics, Speech and Signal Processing (ICASSP)},
  pages     = {436--440},
  year      = {2019},
  publisher = {IEEE},
  doi       = {10.1109/ICASSP.2019.8682574},
}

\end{document}